\documentclass[UKenglish,copyright,creativecommons]{eptcs}
\usepackage{mathtools}
\usepackage{times}
\usepackage{mathpartir,amsmath,amssymb
}
\usepackage{amsmath,amssymb,wasysym,xspace}
\usepackage{mathptmx,stmaryrd}
\usepackage{eurosym}
\usepackage[T1]{fontenc}
\usepackage[mathscr]{euscript}
\usepackage[usenames,dvipsnames]{color}
\usepackage{epsfig}
\usepackage{amsfonts}
\usepackage{latexsym}
\usepackage{breakurl}
\usepackage{listings}
\usepackage{enumerate}
\usepackage{url}
\usepackage{hyperref}
\usepackage{wasysym}
\usepackage{bbm}
\newtheorem{proposition}{Proposition}[section]

\newtheorem{lemma}[proposition]{Lemma}

\newtheorem{definition}[proposition]{Definition}

\newtheorem{theorem}[proposition]{Theorem}

\newcommand{\cinferrule}[3][]{
  \mprset{fraction={===},
  fractionaboveskip=0.2ex,
  fractionbelowskip=0.4ex}
  \inferrule[#1]{#2}{#3}
}
\newcommand{\tin}[3]{#1?#2(#3)}
\newcommand{\tout}[3]{#1!#2(#3)}
\newcommand{\rulename}[1]{\text{\small[\textsc{#1}]}}

\newdimen\proofrulebreadth \proofrulebreadth=.05em
\newdimen\proofdotseparation \proofdotseparation=1.25ex
\newdimen\proofrulebaseline \proofrulebaseline=2ex
\newcount\proofdotnumber \proofdotnumber=3
\let\then\relax
\def\hfi{\hskip0pt plus.0001fil}
\mathchardef\squigto="3A3B
\newif\ifinsideprooftree\insideprooftreefalse
\newif\ifonleftofproofrule\onleftofproofrulefalse
\newif\ifproofdots\proofdotsfalse
\newif\ifdoubleproof\doubleprooffalse
\let\wereinproofbit\relax
\newdimen\shortenproofleft
\newdimen\shortenproofright
\newdimen\proofbelowshift
\newbox\proofabove
\newbox\proofbelow
\newbox\proofrulename
\def\shiftproofbelow{\let\next\relax\afterassignment\setshiftproofbelow\dimen0 }
\def\shiftproofbelowneg{\def\next{\multiply\dimen0 by-1 }%
\afterassignment\setshiftproofbelow\dimen0 }
\def\setshiftproofbelow{\next\proofbelowshift=\dimen0 }
\def\setproofrulebreadth{\proofrulebreadth}

\def\prooftree{
\ifnum  \lastpenalty=1
\then   \unpenalty
\else   \onleftofproofrulefalse
\fi
\ifonleftofproofrule
\else   \ifinsideprooftree
        \then   \hskip.5em plus1fil
        \fi
\fi
\bgroup
\setbox\proofbelow=\hbox{}\setbox\proofrulename=\hbox{}%
\let\justifies\proofover\let\leadsto\proofoverdots\let\Justifies\proofoverdbl
\let\using\proofusing\let\[\prooftree
\ifinsideprooftree\let\]\endprooftree\fi
\proofdotsfalse\doubleprooffalse
\let\thickness\setproofrulebreadth
\let\shiftright\shiftproofbelow \let\shift\shiftproofbelow
\let\shiftleft\shiftproofbelowneg
\let\ifwasinsideprooftree\ifinsideprooftree
\insideprooftreetrue
\setbox\proofabove=\hbox\bgroup$\displaystyle 
\let\wereinproofbit\prooftree
\shortenproofleft=0pt \shortenproofright=0pt \proofbelowshift=0pt
\onleftofproofruletrue\penalty1
}

\def\eproofbit{
\ifx    \wereinproofbit\prooftree
\then   \ifcase \lastpenalty
        \then   \shortenproofright=0pt  
        \or     \unpenalty\hfil         
        \or     \unpenalty\unskip       
        \else   \shortenproofright=0pt  
        \fi
\fi
\global\dimen0=\shortenproofleft
\global\dimen1=\shortenproofright
\global\dimen2=\proofrulebreadth
\global\dimen3=\proofbelowshift
\global\dimen4=\proofdotseparation
\global\count255=\proofdotnumber
$\egroup  
\shortenproofleft=\dimen0
\shortenproofright=\dimen1
\proofrulebreadth=\dimen2
\proofbelowshift=\dimen3
\proofdotseparation=\dimen4
\proofdotnumber=\count255
}

\def\proofover{
\eproofbit 
\setbox\proofbelow=\hbox\bgroup 
\let\wereinproofbit\proofover
$\displaystyle
}%
\def\proofoverdbl{
\eproofbit 
\doubleprooftrue
\setbox\proofbelow=\hbox\bgroup 
\let\wereinproofbit\proofoverdbl
$\displaystyle
}%
\def\proofoverdots{
\eproofbit 
\proofdotstrue
\setbox\proofbelow=\hbox\bgroup 
\let\wereinproofbit\proofoverdots
$\displaystyle
}%
\def\proofusing{
\eproofbit 
\setbox\proofrulename=\hbox\bgroup 
\let\wereinproofbit\proofusing
\kern0.3em$
}

\def\endprooftree{
\eproofbit 
  \dimen5 =0pt
\dimen0=\wd\proofabove \advance\dimen0-\shortenproofleft
\advance\dimen0-\shortenproofright
\dimen1=.5\dimen0 \advance\dimen1-.5\wd\proofbelow
\dimen4=\dimen1
\advance\dimen1\proofbelowshift \advance\dimen4-\proofbelowshift
\ifdim  \dimen1<0pt
\then   \advance\shortenproofleft\dimen1
        \advance\dimen0-\dimen1
        \dimen1=0pt
        \ifdim  \shortenproofleft<0pt
        \then   \setbox\proofabove=\hbox{%
                        \kern-\shortenproofleft\unhbox\proofabove}%
                \shortenproofleft=0pt
        \fi
\fi
\ifdim  \dimen4<0pt
\then   \advance\shortenproofright\dimen4
        \advance\dimen0-\dimen4
        \dimen4=0pt
\fi
\ifdim  \shortenproofright<\wd\proofrulename
\then   \shortenproofright=\wd\proofrulename
\fi
\dimen2=\shortenproofleft \advance\dimen2 by\dimen1
\dimen3=\shortenproofright\advance\dimen3 by\dimen4
\ifproofdots
\then
        \dimen6=\shortenproofleft \advance\dimen6 .5\dimen0
        \setbox1=\vbox to\proofdotseparation{\vss\hbox{$\cdot$}\vss}%
        \setbox0=\hbox{%
                \advance\dimen6-.5\wd1
                \kern\dimen6
                $\vcenter to\proofdotnumber\proofdotseparation
                        {\leaders\box1\vfill}$%
                \unhbox\proofrulename}%
\else   \dimen6=\fontdimen22\the\textfont2 
        \dimen7=\dimen6
        \advance\dimen6by.5\proofrulebreadth
        \advance\dimen7by-.5\proofrulebreadth
        \setbox0=\hbox{%
                \kern\shortenproofleft
                \ifdoubleproof
                \then   \hbox to\dimen0{%
                        $\mathsurround0pt\mathord=\mkern-6mu%
                        \cleaders\hbox{$\mkern-2mu=\mkern-2mu$}\hfill
                        \mkern-6mu\mathord=$}%
                \else   \vrule height\dimen6 depth-\dimen7 width\dimen0
                \fi
                \unhbox\proofrulename}%
        \ht0=\dimen6 \dp0=-\dimen7
\fi
\let\doll\relax
\ifwasinsideprooftree
\then   \let\VBOX\vbox
\else   \ifmmode\else$\let\doll=$\fi
        \let\VBOX\vcenter
\fi
\VBOX   {\baselineskip\proofrulebaseline \lineskip.2ex
        \expandafter\lineskiplimit\ifproofdots0ex\else-0.6ex\fi
        \hbox   spread\dimen5   {\hfi\unhbox\proofabove\hfi}%
        \hbox{\box0}%
        \hbox   {\kern\dimen2 \box\proofbelow}}\doll%
\global\dimen2=\dimen2
\global\dimen3=\dimen3
\egroup 
\ifonleftofproofrule
\then   \shortenproofleft=\dimen2
\fi
\shortenproofright=\dimen3
\onleftofproofrulefalse
\ifinsideprooftree
\then   \hskip.5em plus 1fil \penalty2
\fi
}

\newcommand{\myrule}[3]{\begin{prooftree}
 #1 \justifies   #2   \using{\rln{#3}} \end{prooftree}}
 \newcommand{\myformula}[1]{\\[5pt]\centerline{#1}\\[5pt]}
  \newcommand{\myformulaD}[1]{\\[5pt]\centerline{#1}\\[0pt]}
   \newcommand{\myformulaN}[1]{\\[5pt]\centerline{#1}\\[-7pt]}
  
  \newcommand{\myformulaM}[1]{\\\centerline{#1}}

    \newcommand{\myformulaH}[1]{#1}
  \newenvironment{myitemize}
               {\begin{itemize}\vspace{-2pt}\topsep0pt\parskip0pt\partopsep0pt\itemsep0pt\leftmargin-100pt\itemsep-1pt\labelwidth0pt\labelsep3pt}
               {\vspace{-1pt}\end{itemize}}

\newcommand{\mysection}[1]{\vspace{-5pt}\section{#1}\vspace{-5pt}}
\newcommand{\mysubsection}[1]{\vspace{-5pt}\subsection{#1}\vspace{-5pt}}
\newcommand{\myparagraph}[1]{\vspace{-10pt}\paragraph{#1}\vspace{-5pt}}
 \newenvironment{myenumerate}
               {\begin{enumerate}\vspace{-2pt}
               \topsep0pt\parskip0pt\partopsep0pt\itemsep0pt\leftmargin10pt\itemsep0pt\labelwidth0pt\labelsep3pt}
               {\vspace{-2pt}
               \end{enumerate}}
  \newenvironment{mydefinition}
               {
               \begin{definition}\vspace{-1pt}
               }
               {\end{definition}}
     \newenvironment{mytheorem}
               {
               \begin{theorem}\vspace{-1pt}
               }
               {\end{theorem}}
               \newenvironment{mylemma}
               {
               \begin{lemma}\vspace{-1pt}
               }
               {\end{lemma}}
                \newenvironment{mytable}
               {\begin{table}
               }
               {\end{table}
               }

\newcommand{\E}{{\mathcal E}}

\newcommand{\pc}{~|~}

\newcommand{\red}{\longrightarrow}

\newcommand{\set}[1]{\{#1\}}

\newcommand{\labelx}[1]{\label{#1}}

\newcommand{\lset}{\text{L}}

\newcommand{\wtproc}[4]{\Gamma\vdash #1 \rhd #2\dup #3}
\newcommand{\wtprocG}[5]{#1 \vdash #2 \rhd #3\dup #4}

\newcommand{\Gvti}[4]{\ensuremath{#1\to\pset:\{#2_i({#3}_i). #4_i \}_{i \in I}}}    

\newcommand{\pp}{{\sf p}}
\newcommand{\q}{{\sf q}}
\newcommand{\pr}{{\sf r}}
\newcommand{\G}{\ensuremath{{\sf G}}}
\newcommand{\ST}{\ensuremath{S}}
\newcommand{\mon}{\ensuremath{\sf mon}}
\newcommand{\proc}{\ensuremath{\sf proc}}
\newcommand{\procS}{\ensuremath{{\sf procS}}}

\newcommand{\T}{\ensuremath{\mathsf{T}}}

\renewcommand{\P}{\ensuremath{P}}
\newcommand{\N}{\ensuremath{N}}
\newcommand{\pset}{\ensuremath{\q}}

\newcommand{\ty}{\textbf{t}}
\newcommand{\sig}{\between}
\newcommand{\KK}{\mathcal{K}}
\newcommand{\HH}{\mathcal{A}}
\newcommand{\GG}{\mathbbmss{G}}
\newcommand{\MM}{\mathbbmss{M}}
\newcommand{\PP}{\mathbbmss{P}}
\newcommand{\QQ}{\mathbbmss{Q}}
\newcommand{\End}{\kf{end}}
\newcommand{\next}{\kf{next}}
\newcommand{\true}{\kf{true}}
\newcommand{\false}{\kf{false}}
\newcommand{\new}{\kf{new}}
\newcommand{\inact}{\ensuremath{\mathbf{0}}}
\newcommand{\kf}[1]{\ensuremath{\mathsf{#1}\xspace}}
\newcommand{\Bool}{\kf{Bool}}
\newcommand{\Nat}{\kf{Int}}
\renewcommand{\true}{\kf{true}}
\renewcommand{\false}{\kf{false}}
\newcommand{\e}{\kf{e}}
\newcommand{\x}{x}
\newcommand{\val}{v}
\newcommand{\eval}[2]{#1 \downarrow #2}

\newcommand{\recvar}{\ensuremath{X}}
\renewcommand{\put}{\kf{op}}
\newcommand{\cond}[3]{\kf{if}~ #1 ~\kf{then} ~#2 ~\kf{else}~#3}
\newcommand{\sep}{\ensuremath{~\mathbf{|\!\!|}~ }}
\newcommand{\fun}{F}

\renewcommand{\l}{\lambda}
\newcommand{\M}{{\sf M}}

\newcommand{\sendMg}[2]{{#1}!\set{\ell_i (#2_i). \M_i}_{i \in I}}
\newcommand{\recMg}[2]{{#1}?\set{\ell_i (#2_i). \M_i}_{i \in I}}

\newcommand{\ch}{{\mathsf{s}}}
\newcommand{\cc}{{\mathsf{c}}}

\newcommand{\stackred}[1]{\xrightarrow{#1}}
\newcommand{\proj}[2]{#1 \!  \upharpoonright  \! #2\,}
\newcommand{\rest}[1]{(\nu\,{#1})}
\newcommand{\procset}{\mathcal{P}}
\newcommand{\NS}{\mathcal{S}}
\newcommand{\pcn}{~|\!|~}
\newcommand{\Q}{\ensuremath{Q}}
\newcommand{\sub}[2]{\set{#1/#2}}

\newcommand{\leqt}{\leq}

\newcommand{\Tset}{\mathcal{T}}
\newcommand{\TT}{\mathbb{T}}
\newcommand{\cuni}{\Cup}   

\newcommand{\typout}[4]{ ! #2(#3). #4}
\newcommand{\typin}[4]{ ? #2(#3).#4}

\newcommand{\procin}[4]{#1 ?  #3 .#4}
\newcommand{\procout}[4]{#1 ! #3. #4}

\newcommand{\agree}{\propto}
\newcommand{\mt}[1]{|#1|}

\newcommand{\pt}{\kf{pa}}
\newcommand{\rln}[1]{\textsc{#1}}
\newcommand{\dup}{ :}

\newcommand{\IF}{\textcolor{blue}{\kf{iF}}}
\newcommand{\AF}{\textcolor{Mahogany}{\kf{aF}}}
\newcommand{\IS}{\textcolor{blue}{\kf{iS}}}
\newcommand{\Ro}{\textcolor{blue}{\kf{Ro}}}
\newcommand{\Lo}{\textcolor{blue}{\kf{Lo}}}
\newcommand{\AS}{\textcolor{Mahogany}{\kf{aS}}}
\newcommand{\NY}{\textcolor{Mahogany}{\kf{NY}}}
\newcommand{\Ch}{\textcolor{Mahogany}{\kf{Ch}}}

\newcommand{\GM}{\textcolor{ForestGreen}{\kf{GM}}}

\newcommand{\FF}{\textcolor{red}{\kf{FF}}}
\newcommand{\TS}{\textcolor{red}{\kf{TS}}}

\DeclareMathAlphabet{\mathpzc}{OT1}{pzc}{m}{it}

\definecolor{ocre}{rgb}{.92,.86,.3}%
\definecolor{verde}{rgb}{0.50,0.70,0.4}
\definecolor{beppeblue}{rgb}{0,0,0.4}
\definecolor{lucagreen}{rgb}{0,0.3,0}
\definecolor{lightgreen}{rgb}{0.79,0.86,.79}
\definecolor{lesslightgreen}{rgb}{0.39,.9,.39}
\definecolor{mydarkgreen}{rgb}{0.10,0.43,0}
\definecolor{darkred}{rgb}{0.67,0.27,0.39}
\definecolor{ultralightgray}{gray}{0.85}
\definecolor{midgray}{gray}{0.6}
\definecolor{siena}{rgb}{0.58,0.23,0.34}
\definecolor{mgreen}{rgb}{0,0.5,0}
\definecolor{lucared}{rgb}{0.8,0,0}

\newcommand{\pskip}[1]{}

\newcommand{\CC}{\mathcal{C}}

\newcommand{\del}[1]{}

\providecommand{\event}{PLACES 2016}
\title{Parallel Monitors for Self-adaptive Sessions\thanks{Partly supported
    by the COST Action IC1201 BETTY.}}
\author{Mario Coppo\thanks{Partly supported by EU H2020-644235 Rephrase project, EU H2020-644298 HyVar project, ICT COST Actions IC1402 ARVI and Ateneo/CSP project RunVar.}
\institute{
Universit\`a di Torino,
coppo@di.unito.it
} \and Mariangiola Dezani-Ciancaglini\thanks{Partly supported by EU H2020-644235 Rephrase project, EU H2020-644298 HyVar project, ICT COST Actions IC1402 ARVI and Ateneo/CSP project RunVar.}
\institute{
Universit\`a di Torino,
dezani@di.unito.it}
\and
Betti  Venneri \institute{
Universit\`a di
Firenze,
venneri@unifi.it}
}

\def\titlerunning{Parallel Monitors for Adaptive Sessions}
\def\authorrunning{Coppo, Dezani-Ciancaglini, \& Venneri }

\begin{document}

\maketitle

\begin{abstract}
%
%

 The paper presents a data-driven  model  of self-adaptivity for multiparty sessions. System choreography is prescribed by a global type. Participants  are incarnated by processes associated with monitors, which control their behaviour. Each participant can access and modify a set of global data, which are able to trigger adaptations in the presence of critical changes of values.\\
The use of the parallel composition for building global types, monitors and processes enables a significant degree of flexibility: an adaptation step can dynamically reconfigure a set of participants only, without altering the remaining participants, even if the two groups communicate.


\end{abstract}

\mysection{Introduction}
The topic of self-adaptiveness has become 
a key research subject within various application domains, as a  response to the growing complexity of software systems operating in many different scenarios and in highly dynamic environments. To manage this complexity at a reasonable cost, novel approaches are needed  in which a system can promptly react to crucial changes  by reconfiguring its  behaviour autonomously and dynamically, in accordance with evolving policies and objectives.

This paper proposes a \emph{data-driven} model of  \emph{self-adaptivity} into the formal framework of
\emph{multiparty sessions}~\cite{CHY08}. Each participant can access and modify the \emph{control data}, whose critical values are responsible for throwing
adaptation steps, in such a way that the system responds to dynamic changes by reconfiguring itself.
 A system comprises four active parties, as in \cite{CDV15}: \emph{global types}, \emph{monitors}, \emph{processes}, and \emph{adaptation functions}. Differently, the proactive role of control data in triggering adaptation is the hallmark of our approach.

A global type represents the overall communication choreography~\cite{CHY12};
its projections onto participants generate the monitors, which
set-up the communication protocols of the participants. The association of  a monitor with a compliant process, dubbed {\em monitored process}, incarnates a participant,  where the process provides the implementation of the monitoring protocol.  Note that
global types, monitors and processes are built using parallel composition. This is the main novelty of our calculus, which  provides the system with rather flexible adaptation capabilities. Using monitors in parallel,  the system is able to minimise self-adaptation by reconfiguring only some of the communications and leaving the other ones as before, in case only a 
part of the whole behaviour needs to be changed. Thus the adaptation of a group of participants can be transparent to another group, also in presence of communications between them, while new participants can be added and/or  some of the old participants can be removed.

The adaptation strategy is owned by control data and adaptation functions. The adaptation function contains the dynamic evolution policy, since it prescribes  how the system needs to reconfigure itself, based on the changes of the data value. As long as control data do not yield critical values, the adaptation function is undefined. Thus participants communicate and read/write data according to their monitors and codes, without performing any reconfiguration.

To exemplify our approach, let us consider a company including an Italian factory ($\IF$), an  Italian supplier ($\IS$) and a store ($\Ro$) in Rome. Control data hold
information about factories, suppliers and stores.
The supplier interacts in parallel with the factory and the store about marketing data (number of item requested, delivery date).
    The interactions are described by the following (single threaded) global types:
\myformulaM{$\G_1=\mu\ty. \IS\to\IF: SF(\text{Item,Amount}). \IF\to\IS: FS(\text{DeliveryDate}).\ty$}
\myformulaM{$\G_2=\mu\ty. \Ro\to\IS: RS(\text{Item,Amount}). \IS\to\Ro: SR(\text{DeliveryDate}).\ty$}
The global type of the system is then given by $\GG_0 = \G_1 \pc \G_2$.

The monitors 
of the participants are obtained as the parallel composition of the projections of these global types. For instance the monitor of participant $\IS$ is:
\myformulaM{ $ \mu\ty.\IF! SF(\text{Item, Amount}). \IF ? FS(\text{Date}). \ty  \pc  \mu\ty.\Ro ? RS(\text{Item, Amount}). \Ro ! SR(\text{Date}). \ty$
}
and a possible process code is  \footnote{Sorts are written with upper case initials, expression $variables$ and $\kf{values}$ with lower case initials, but different fonts.}:
 \myformulaM{ $ \mu X.y! SF({\kf{item, amount}}). y ? FS(date). X  \pc  \mu X.y ? RS(item, amount). y ! SR({\kf{date}}). X$
}
where $!$ represents output, $?$ represents input, and $y$ is a channel.

Assume that the control data reveals the opening of a new store ($\Lo$) in London, which must receive its items from the Italian seller.

A new global type is obtained by putting in parallel to the current global type the type $\G$, produced by applying the adaptation function to the 
control data:
\myformulaM{$\G =\mu\ty. \Lo\to\IS: LS(\text{Item,Amount}). \IS\to\Lo: SL(\text{DeliveryDate}).\ty$}
The global type $\G$ adds the participant $\Lo$ to the conversation and modifies the participant $\IS$ by adding the monitor
\myformulaH{$\mu\ty.\Lo ? LS(\text{Item, Amount}). \Lo ! SL(\text{Date}). \ty$
and the
process
\myformulaM{$
\mu X.y ? LS(item, amount). y ! SL({\kf{date}}). X$
}
}
in parallel with his previous ones.
%

\myparagraph{Outline} This paper has a standard structure. After a further example (Section~\ref{me}) we present syntax (Section~\ref{syntaxSection}), types
(Section~\ref{ts}) and semantics (Section~\ref{semanticSection}) of our calculus. In Section~\ref{crw} we draw some conclusions and discuss related works.

\mysection{An Extended Example}\labelx{me}
Let us now extend the example of the Introduction by considering a company Ada in which
 factories interact with the general manager about production policies and suppliers interact with factories and stores by exchanging marketing data. Moreover, the stores can communicate each 
 other for requiring some products.
The company initially consists of:
\begin{myitemize}
\item a general manager (\GM);
\item an Italian (\IF) and an American Factory (\AF);
\item  an Italian (\IS) and an American Supplier (\AS);
\item three stores, located in Rome (\Ro), New York (\NY) and Chicago (\Ch).
\end{myitemize}
The global type prescribing the communications  is $\GG=\G_1\pc\G_2\pc\G_3\pc\G_4\pc\G_5\pc\G_6$, where $\G_1, \G_2$ are as in the Introduction and:
\myformulaM{$\G_3=\mu\ty. \GM\to\IF: GIF(\text{ProductionLines}). \GM\to\AF: GAF(\text{ProductionLines}).$}
 \myformulaM{\hspace{1.4cm}$\IF\to\GM: IFG(\text{ProgressReport}). \AF\to\GM: AFG(\text{ProgressReport}).\ty$}
\myformulaM{$\G_4=\mu\ty. \AS\to\AF: SF(\text{Item,Amount}). \AF\to\AS: FS(\text{DeliveryDate}).\ty$}
\myformulaM{$\G_5=\mu\ty. \Ch\to\AS: CS(\text{Item,Amount}). \AS\to\Ch: SC(\text{DeliveryDate}).\ty$}
\myformulaM{$\begin{array}{rcl}
\G_6 & = & \mu\ty. \NY\to\AS: NS(\text{Item,Amount}). \\
 & & \phantom{\mu\ty.}\AS\to\NY:\{YES(\text{DeliveryDate}).\NY\to\Ch: NCY(\text{NoItem}).\ty, \\
 &  & \phantom{\mu\ty.\NY\to }   
 NO(\text{NoItem}): \NY\to\Ch: NCN(\text{Item,Amount}).
  \Ch\to\NY: CN(\text{DeliveryDate}).\ty \} \end{array} $}
Note that the New York store can ask for items both to the America factory and to the Chicago store.
Notice also  that $\G_1$ and $\G_4$ differ for the participants, but not for the labels, and this allows the process
\myformulaM{$\mu X. y?SF(item,amount).y!FS(\kf{deliveryDate}).X$}
to incarnate both $\IF$ for $\G_1$ and $\AF$ for $\G_4$. This process in parallel with \myformulaM{$\mu X. y?GIF(\text{ProductionLines}).y!IFG(\text{ProgressReport}).X$,}
where $y$ must be replaced  by the appropriate run time channel, plays the role of
$\IF$ for the whole $\GG$.


Assume now that the catastrophic event of a fire, incapacitating the American factory, requires the Ada company to update itself. The American factory modifies the control data by putting the information about its unavailability.
The function $F$, applied to the modified control data, kills $\AF$ and  produces the global type:
\myformulaM{$\begin{array}{lll}\GG'&=&\FF\to\GM: FG(\text{DamagesReport}). \\ & & \mu\ty.\GM\to\TS: GT(\text{ReconstructionPlan}).\TS\to\GM: TG(\text{TechnicalRevisions}).\ty \pc \\
 & & \mu\ty. \AS\to\IF: ASF(\text{Item,Amount}). \IF\to\AS: FAS(\text{DeliveryDate}).\ty\pc\\
 &&\mu\ty. \GM\to\IF: GIF(\text{ProductionLines}).\IF\to\GM: IFG(\text{ProgressReport}).\ty \end{array}$}
According to $\GG'$, the FireFighters ($\FF$) send to the general manager a report on the damages and then the general manager opens with the Technical Service ($\TS$) a conversation about the reconstruction plan.   
The American seller now asks for products to the Italian factory. Notice that the labels of the communications between them must be different from those in $\G_1$, which is not modified, since the Italian factory continues to serve the Italian seller as before. The new global type is the parallel of $\GG'$ with  the global type obtained from $\GG$ by
erasing $\G_3$ (in which all communications involve $\AF$ or they are between $\GM$ and $\IF$) and
erasing $\G_4$ (in which all communications involve $\AF$). The rational is to erase the communications in which
\begin{myitemize} \item either one of the two participants is killed,
\item or both participants are reconfigured.
 \end{myitemize}
The new process incarnating $\IF$ is obtained by putting
{\small\myformulaM{$\mu X. y?ASF(\text{item,amount}).y!FAS(\text{deliveryDate}).X\pc\\
\mu X. y?GIF(\text{ProductionLines}).y!IFG(\text{ProgressReport}).X$}}
 in parallel with the previous process incarnating $\IF$. Then the code of the $\IF$ participant becomes:
 \myformulaM{$\begin{array}{c}\mu X. y?SF(item,amount).y!FS(\kf{deliveryDate}).X \pc
 \mu X. y?ASF(item,amount).y!FAS(\kf{deliveryDate}).X\pc\\
 \mu X. y?GIF(\text{ProductionLines}).y!IFG(\text{ProgressReport}).X
 \end{array}$}
where $y$ must be replaced by the appropriate run time channel.

For readability in this example (and in that of the Introduction) we described adaptation as modification of global types. In the semantics (Section~\ref{semanticSection}) global types are left implicit and the adaptation rule modifies monitors and processes.

\mysection{Syntax}  \labelx{syntaxSection}

 \subsection{Global Types}       

 Following a widely common approach~\cite{CHY12}, the set-up of protocols starts from global types. Global types establish overall communication  schemes.
In our setting they also control the reconfiguration phase, in which a system adapts itself to new environmental conditions.

Let $\lset$ be a set of \emph{labels}, ranged over by $\ell$, which mark the exchanged values as in~\cite{DY11}.
We assume some basic \emph{sorts}, ranged over by $S$, i.e.
 $S$  ::=  $\Bool~\sep~\Nat~\sep~\ldots$.

Single threaded global types give sequential communications with alternatives.\\ In  $\Gvti \pp \ell \ST {\G}$ participant $\pp$ sends to participant $\q$ a label $l_i$ together with a value of sort $S_i$ for some $i\in I$. We implicitly assume that $\ell_i \neq \ell_j$  for all $i\neq j$. \\
We take an equi-recursive view of global types, monitors, processes and process types, identifying $\mu\ty.\G$ with $\G\sub{\mu\ty.\G}\ty$ etc. We assume that all recursions are guarded. In writing examples we omit brackets when there is only one branch.

Global types are parallel compositions of single threaded global types with distinct labels for common participants.
\begin{mydefinition}\labelx{sgtypesdef}
\begin{myenumerate}
\item {\em Single threaded global types} are defined by:
\myformulaN{
 $\G~~ ::= ~~\Gvti \pp \ell \ST {\G} ~\sep ~\mu\ty.\G ~\sep ~\ty ~\sep ~\End$
}
\item {\em Global types} are defined by:
\myformulaH{
 $\GG~~ ::= ~~\G~\sep \GG\pc\GG$
},
where global types in parallel have distinct labels for common participants.
\end{myenumerate}
\end{mydefinition}
We allow parallels of global types with shared participants, a possibility usually forbidden~\cite{CHY08,CHY12,BCDHY13}.

\mysubsection{Monitors}                          
Single threaded monitors can be viewed as projections of single threaded global types onto individual participants, as in the standard approach of~\cite{CHY08} and~\cite{BCDDDY08}. In our calculus, however, monitors are more than types, as in~\cite{CDV15}: they have an active role in system dynamics, since they guide communications and adaptations. As expected monitors are parallel compositions of single threaded monitors with distinct labels.

\begin{mytable} {
\myformulaH{
$\proj{(\Gvti\pp\ell \ST {\G})}\pr=\begin{cases}
   \pp ? \set{\ell_i(\ST_i).\proj{\G_i}\q}_{i \in I} & \text{if }\pr=\q \\
    \pset!\set{\ell_i(\ST_i).\proj{\G_i}\pr}_{i\in I}  & \text{if }\pr=\pp\\
   \proj{\G_{i_0}}{\pr} & \text{where $i_0\in I$ if $\pr \neq \pp$ and $\pr \neq\q$} \\
                                      & \text{and $\proj{\G_i}{\pr} = \proj{\G_j}{\pr}$ for all $i,j\in I$} \\
   {\sf u} ? \bigcup_{k \in K}  \set {\ell_k (\ST_k) . \proj{\G_k}{\pr}} &
                                    \text{where $K =\bigcup_{i\in I} J_i$ if $\pr \neq \pp$,  $\pr \neq\q$  and }\\
                                    &  \proj{\G_{i}}{\pr} =  {\sf u} ? \set{\ell_j(\ST_j).\G_j}_{j \in J_i} \text{ for all $i\in I$ and}
                                   \\ & J_m \cap J_n = \emptyset
                               \text{ and }\ell_m \neq \ell_n
                               \\&\text{for all }m,n \in K \text{ such that } m \neq n

\end{cases}$
\\[3mm]
 $\proj{(\mu\ty.\G)}\pp=\begin{cases}
  \mu\ty. \proj{\G}{\pp}   & \text{if }\pp\in\G, \\
   \End   & \text{otherwise}.
\end{cases}$~~~~~$\proj{\ty}{\pp} = \ty $~~~~~~~$\proj{\End}{\pp} = \End $ ~~~~~~~~~~~~ $\proj{(\GG\pc\GG')}{\pp} = (\proj{\GG}{\pp})\pc(\proj{\GG'}{\pp}) $
}
}
\caption{ Projection of a global type onto a participant.}\labelx{projection}
\end{mytable}

\begin{mydefinition}\labelx{smonitorsdef}
\begin{myenumerate}
\item \emph{Single threaded monitors} are defined by:
\myformulaN{
$ \M ~~  ::= ~~ \recMg{\pp}{S} ~  \sep  ~ \sendMg{\pset}{S} ~\sep
 ~
    \mu\ty.\M   ~\sep  ~ \ty   ~\sep ~  \End $
}

\item \emph{Monitors} are defined by:
\myformulaH{
 $\MM ~~  ::= ~~ \M  ~ \sep ~   \MM\pc\MM
$
},
where monitors in parallel have distinct labels.
\end{myenumerate}
\end{mydefinition}  
An input monitor $\recMg{\pp}{S}$  is able to drive a process that can receive, for each $i\in I$, a value of sort $\ST_i$, labeled by $\ell_i$, having as continuation a process which is adequate for $\M_i$. This corresponds to an external choice. Dually an output monitor $\sendMg{\pset}{S}$ is able to drive  a process which can send (by an internal choice) a value of sort $\ST_i$, labeled by $\ell_i$, and
then continues as prescribed by $\M_i$  for each $i\in I$.

The projection of global types onto participants is given in Table~\ref{projection}. A projection onto a participant $\pr$ not involved, as sender or receiver, in a choice is defined if either the projection onto $\pr$ of all continuations are the same (condition $\proj{\G_i}{\q} = \proj{\G_j}{\q}$ for all $i,j\in I$) or all these projections can be merged in an input monitor (last case), as in~\cite{DY11}. A global type $\G$ is {\em well formed} if its projections are defined  for all participants. We assume always that global types are well formed.

\mysubsection{Processes}        
Processes represent code that is associated to monitors in order to implement participants. As in~\cite{CDV15} and
differently from standard session calculi (see~\cite{wg1soar} and the references there), processes do not specify the participants involved in sending and receiving actions. The associated monitors determine senders and receivers.

The communication actions of processes are performed through \emph{channels}.  Each process owns a unique channel endpoint. We use $y$ to denote this channel endpoint  in the process code.
 As usual,
 the user channel $y$ will be replaced at run time by a session channel $\ch[\pp]$ (where $\ch$ is the session name and $\pp$ is 
 the current participant). Let $\cc$ denote a user channel or a session channel.
\begin{mydefinition}\labelx{sprocdefinition}\begin{myenumerate}
\item \emph{Single threaded processes} are defined by:
\myformulaN{\hspace{-20pt}
\begin{tabular}{rcl}
 $\P $ & ::=  & $\procin  \cc  \pp {\ell(\x)} \P   ~~\sep~~  \procout \cc \pset {\ell(\e)} \P ~~\sep~~  \P +\P  ~~\sep~~ \cond{\e}{\P}{\P}
  ~~\sep~~ \put. \P ~~\sep~~   \mu\recvar.\P ~~\sep~~  \recvar ~~\sep~~\inact$
    \end{tabular}
}
\item
\emph{Processes} are defined by:
\myformulaH{
\begin{tabular}{rclll}
 $\PP $ & ::= & \P & \sep $\PP\pc\PP$.
    \end{tabular}
}
\end{myenumerate}
\end{mydefinition}
The syntax of processes is rather standard, in particular the operator $+$ represents external choice.  A system has control data (see Definition~\ref{sysdef}) and the $\put$ operator represents an action on this data, for instance a ``read'' or ``write'' operation. We leave unspecified the kind of actions, since we are only interested in the dynamic changes of this data, which determine the self-reconfiguration of the whole system.

\mysubsection{Networks}     
A process is always controlled by a monitor, which ensures that all performed actions fit the protocol prescribed by the global type. Each monitor controls a single process. So participants correspond to pairs of processes and monitors. We write $\MM[\PP]$ to represent a process $\PP$ controlled by a monitor $\MM$, dubbed {\em monitored process}.
The sessions are initiated by the ``$\new$'' constructor applied to a global type ({\em session initiator}), denoted by $\new(\GG)$, which generates the monitors and associates them with adequate processes (see Definition~\ref{ade}).
The parallel composition of session initiators and processes with the corresponding monitors form a network. Networks can be restricted on session names.
\begin{mydefinition}\labelx{netdef}
 {\em Networks} are defined by:
\myformulaH{ $\N ~ ::= ~\new(\GG) ~\sep~\MM[\PP] ~\sep \N\pcn\N  ~\sep~ (\nu\ch)\N$.}\end{mydefinition}
By $\MM_\pp[\PP_\pp]$ we denote the monitored process of participant $\pp$, i.e. the channel in $\PP_\pp$ must be $\ch[\pp]$ for some $\ch$. We use $\Pi$ to represent the parallel composition of monitored process.

\mysubsection{Systems}       
A system includes a network, a global state and assumes a collection of processes denoted by $\procset$. A main component of the global state is the {\em adaptation function} $F$, which says how to run adaptations according to critical data variations.
The global state contains also {\em control data} $\sigma$, which are updated by processes.
We represent systems as the composition (via ``$\sig$'') of a network and control data. We omit the adaptation function and the process collection,
considering them as implicit parameters, since they are not modified by reductions.
\begin{mydefinition}\labelx{sysdef}
 {\em Systems} are defined by:
\myformulaH{ $ \NS ~ ::= ~ \N \sig \sigma$.}\end{mydefinition}

\mysection{Type System}\labelx{ts}

Process types describe process communication behaviours~\cite{HKV98}. They have prefixes corresponding to sending and receiving of labels and values.

\begin{mydefinition}\begin{myenumerate}
\item
\emph{Single threaded types} are inductively defined by:
\myformula{$\T ~  ::=  ~ \bigwedge_{i\in I}\typin{\pp}{\ell_i} {\ST_i} {\T_i}  \sep
      \bigvee_{i\in I}\typout \pset {\ell_i} {\ST_i} {\T_i}  \sep
        \mu\ty.\T\sep \ty\sep\End$}
        where all labels in intersections and unions are different.\vspace{2pt}
\item
\emph{Types} are inductively defined by:
\myformulaH{$\TT ~  ::=  ~  \T   \sep
         \TT\pc\TT  $},
         where types in parallel have distinct labels.
\end{myenumerate}
\end{mydefinition}

Types are built from input and output prefixes ($?l(S)$
and $!l(S)$) by means of constructs for intersection types
(used to type external choices)
and union types (used to type conditionals).

\begin{mytable}
\myformulaH{$
\begin{array}{c}
\wtproc{\inact}\cc{\End}{\Delta} ~~\rln{end}\quad
 \myrule{\wtproc{\P}\cc{\T}{\Delta}}{\wtproc{\put.\P}\cc{\T}{\Delta}}{op}\quad
 \wtprocG{\Gamma,\recvar:\T} {\recvar} {\cc}{\T}{} ~~\rln{ax} \\ \\
\myrule{\wtprocG{\Gamma,\recvar:\T} \P \cc \T \Delta}
 {\wtproc {\mu\recvar. \P}{\cc}{\T}{\Delta\sub{\T}{\recvar}}}{rec}\qquad
 \myrule {\wtprocG{\Gamma,x:\ST}{\P}{\cc}{\T}{\Delta}}
{\wtproc{\procin \cc \pp {\ell(\x)} \P}{\cc}{\typin \pp \ell \ST  \T} {\Delta}}{rcv}\qquad
 \myrule{\wtproc{\P} \cc {\T} {\Delta}~~~~~~~~~\Gamma\vdash \e:\ST}
{\wtproc{\procout \cc \pset{\ell(\e)} \P}{\cc}{\typout\pset \ell \ST \T}{\Delta}}{send}
\\ \\
  \myrule{\wtproc{\P_1}{\cc}{\T_1}{\Delta_1}\quad\wtproc{\P_2}{\cc}{\T_2}{\Delta_2}\quad\T_1 \wedge \T_2\in \Tset}
 {\wtproc{\P_1 + \P_2}{\cc}{\T_1 \wedge \T_2}{\Delta_1 \cuni \Delta_2}}{choice}\\\\
 \myrule{\Gamma \vdash \e : \Bool\quad\wtproc{\P_1}{\cc}{\T_1}{\Delta_1}\quad\wtproc{\P_2}{\cc}{\T_2}{\Delta_2}\quad\T_1 \vee \T_2\in \Tset}
 {\wtproc{\cond \e {\P_1}{\P_2}}{\cc}{\T_1 \vee \T_2}{\Delta_1 \cuni \Delta_2}}{if} \\\\
 \myrule{\wtproc{\PP_1}{\cc}{\TT_1}{\Delta_1}\quad\wtproc{\PP_2}{\cc}{\TT_2}{\Delta_2}}
 {\wtproc{\PP_1 \pc \PP_2}{\cc}{\TT_1 \pc \TT_2}{\Delta_1 \cuni \Delta_2}}{par}
 \end{array}
 $}
\caption{ Typing rules for processes.}\labelx{tr}
\end{mytable}

An \emph{environment} $\Gamma$ is a finite mapping from expression variables to sorts and from process variables to single threaded  types:
 \myformulaH{$\Gamma::=\emptyset~\sep~\Gamma,\x\dup S~\sep~\Gamma,\recvar \dup \T$}, where the notation $\Gamma, x\dup S$ ($\Gamma, \recvar\dup \T$) means that $\x$ ($X$) does not occur in $\Gamma$.

We assume that expressions are typed by sorts, as usual. The typing judgments for expressions are of the shape
\myformulaH{$\Gamma\vdash \e:\ST$}
and the typing rules for expressions are standard.

Single threaded processes can be typed only if they contain at most one channel. This choice is justified by the design of monitored processes as session participants and by the absence of delegation. Therefore the typing judgments for single threaded processes have the form
\myformulaH{$\wtproc{\P}{\cc}{\T}{\Delta}$} and for processes
\myformulaH{$\wtproc{\PP}{\cc}{\TT}{\Delta}$}.
Typing rules for processes are given in Table~\ref{tr}. The external choice is typed by an intersection type, since an external choice offers both behaviours of the composing processes. Dually, a conditional is an internal choice and so it is typed by a union type.

As in~\cite{CDV15}, a single threaded process is equipped with a single threaded type, ranged over
by $\T$, describing its communication actions. A parallel composition of single threaded processes is typed by the parallel composition of
the corresponding single threaded types. We use $\TT$ to range over types and we write $\wtprocG{}{\PP}{\cc}{\TT}{}$ if process $\PP$ has the unique channel $\cc$ typed by $\TT$.

\begin{mytable}
\centerline{$
\begin{array}{@{}c@{}}
\inferrule[\rulename{sub-end}]{}
  {\End \leqt \End}\qquad
\cinferrule[\rulename{sub-in}]{
\forall i\in I: \quad \T_i \leqt\T_i' }{
  \bigwedge_{i\in I\cup J} \tin\pp{\ell_i}{\ST_i}.\T_i \leqt \bigwedge_{i\in I}\tin\pp{\ell_i}{\ST_i}.\T_i'
}
\qquad
\cinferrule[\rulename{sub-out}]{
 \forall i\in I: \quad  \T_i \leqt \T'_i
}{\bigvee_{i\in I}
  \tout\pp{\ell_i}{\ST_i}.\T_i
  \leqt\bigvee_{i\in I\cup J}
 \tout\pp{\ell_i}{\ST_i}.\T_i'
 }
\end{array}
$}
\caption{Subtyping.}\labelx{tam}
\end{mytable}

We end this section defining {\em adequacy}
between a process $\PP$ and a monitor $\MM$,
denoted $\PP\agree\MM$, which is based on the matching between types and monitors.
This matching 
is made rather flexible by using the {\em subtype} relation on types defined in Table~\ref{tam}. The double line in rules indicates that the rules
 are interpreted {\em
   coinductively}~\cite[21.1]{PierceBC:typsysfpl}. Subtyping is monotone, for input/output prefixes, with respect to continuations and it follows the usual set theoretic inclusion of intersection and union.
\begin{mydefinition}\labelx{ade}\begin{myenumerate}
\item A single threaded  process $\P$  is {\em adequate} for a single threaded monitor $\M$  (notation $\P\agree\M$) if  $\wtprocG{}{\P}{\cc}{\T}{}$ and $\T\leqt \mt\M$, where the mapping $\mt{\;}$ is defined by:
\myformulaD{$\begin{array}{c}
\mt{\recMg{\pp}{S}}=\bigwedge_{i \in I} \typin {\pp} {\ell_i} {\ST_i}{\mt{\M_i}}\qquad
\mt{\sendMg{\pset}{\ST}}=\bigvee_{i \in I} \typout {\pset} {\ell_i} {\ST_i}{\mt{\M_i}}\\[.6em]
\qquad \mt {\mu\ty.\M} =\mu\ty.\mt\M \qquad\mt\ty=\ty\qquad
\mt\End=\End
\end{array}$}\item
The {\em adequacy}  of a process $\PP$ for a  monitor $\MM$  (notation $\PP\agree\MM$) is the smallest relation such that:
\begin{myitemize} \item $\P\agree\M$;
\item $\PP_1\agree\MM_1$ and $\PP_2\agree\MM_2$ imply $\PP_1\pc\PP_2\agree\MM_1\pc\MM_2$.
\end{myitemize}
\end{myenumerate}
\end{mydefinition}\vspace{3pt}
Note that adequacy is decidable, since processes have unique types and subtyping is decidable.

 The following lemma states a crucial property of the adequacy relation. A process $\PP$ is adequate for a monitor $\MM$ iff $\PP$ is the parallel composition of single threaded processes, $\MM$ is the parallel composition of single threaded monitors and each single threaded process is adequate for exactly one single threaded monitor and vice versa.
\begin{mylemma}\labelx{la}
If $\PP\agree\MM$, then $\PP=\Pi_{i\in I} \P$ and $\MM=\Pi_{i\in I} \M$ such that $\P_i\agree\M_i$ for all $i\in I$ and $\P_i\not\agree\M_j$ for all $i,j\in I$ with $i\not=j$.
\end{mylemma}
Under the hypothesis of previous lemma we use $\PP.i$ and $\MM.i$ to denote $\P_i$ and $\M_i$, respectively.

\mysection{Semantics}\labelx{semanticSection}

Processes can communicate labels and values, or can read/modify the control data trough $\put$ operations. The semantics of processes is described via a LTS defined in Table~\ref{ltsp}, where the treatment of recursions and conditionals is standard.

 \begin{mytable}[b]
 \normalsize
\centerline{$\begin{array}{c}
\put.\P \stackred{\put} \P \qquad \mu\recvar.\P \stackred{\tau} \P\sub{\mu\recvar.\P}{\recvar} \qquad 
\procin {\ch[\pp]} \q {\ell(\x)} \P \stackred{\ch[\pp] ? \ell (\val)} \P \sub{\val}{\x}  \qquad \procout{\ch[\pp]} \pset {\ell(\e)} \P \stackred{\ch[\pp] ! \ell(\val)} \P ~~ \eval{\e}{\val}\\[.4em]
$\cond{\e}{\P}{\Q}$  \stackred{\tau} \P ~~~\eval{\e}{\true}
 \qquad \qquad
$\cond{\e}{\P}{\Q}$  \stackred{\tau} \Q ~~~\eval{\e}{\false} \\[.5em]
\myrule{\P \stackred{\beta} \P'}{\P + \Q \stackred{\beta} \P'}{} \qquad  \myrule{\P \stackred{\gamma} \P'}{\P +\Q \stackred{\gamma} \P'+\Q}{}
\qquad \myrule{\PP\stackred{\delta}\PP'}{\PP\pc\PP''\stackred{\delta}\PP'\pc\PP''}{}
\end{array}$}
\caption{\normalsize LTS of processes.}\labelx{ltsp}
\end{mytable}

 In the rules for external choice, $\beta$ ranges over  $\ch[\pp] ? \ell (\val), \ch[\pp] ! \ell(\val)$ and $\gamma$ ranges over $\put,\,\tau$. In the rule for parallel composition $\delta$ ranges over $\beta$ and $\gamma$. We omit the symmetric rules. The external choices are done by the communication actions, while the operations on the global state are transparent. This is needed since the operations on the control data are recorded neither in the process types nor in the monitors. An operation on data in an external choice can be performed also if a branch, different from that containing the operation, is then executed.

We assume a standard structural equivalence on processes, monitors, and networks, in which the parallel operator is commutative and associative, and $\inact$ and $\End$ are neutral elements. Moreover,  the process $\inact$ with $\End$ monitor behaves as the neutral element for the parallel of networks and it  absorbs restrictions, that is:
 \myformulaH{$\End[\inact]\pc\N~\equiv~\N$  ~~~\text{and} ~~~$(\nu\ch)\End[\inact]~\equiv~\End[\inact]$.}

The evolution of a system depends on the evolution of its network and control data. The basic components of networks are the openings of sessions (through the \new\ operator on global types) and the  processes associated with monitors.

A session starts by reducing a network $\new(\GG)$. If $\GG$ is a single threaded global type, then for each $\pp$ in the set $\pt(\G)$ of the participants in the global type $\G$, we need to find a process $\P_\pp$ in the collection $\procset$ associated to the current system. The process $\P_\pp$ must be  adequate for the monitor which is the projection of $\G$ onto $\pp$. Then the process (where the channel $y$ has been replaced by $s[\pp]$) is associated to the corresponding monitor. Lastly, the name $\ch$ is restricted.
\myformulaD{$\myrule{ \M_\pp = \proj \G \pp~~~~\forall \pp \in \pt(\G).~ \P_\pp \in \procset ~\&~ \P_\pp\agree\M _ \pp}
{\new(\G)  ~ \red{} \rest{\ch}~ (\prod_{\pp \in \pt(\G)} \M_\pp [\P_\pp\sub{\ch[\pp]}y]) }{InitS}$}
The $\new$ applied to the parallel of two global types simply puts in parallel the networks produced by the $\new$s of the two global types, by restricting them with the same session name.
In this way we ensure the privacy of the communications in a session (as standard in session calculi~\cite{CHY08}).
\myformulaD{$\myrule{\new(\GG)  ~ \red{} \rest{\ch}~\N\quad \new(\GG')  ~ \red{} \rest{\ch}~\N'}
{\new(\GG\pc\GG')  ~ \red{} \rest{\ch}~(\N\pc\N') }{InitM}$}

Monitors guide the communications of processes by choosing the senders/receivers and by allowing only some actions among those offered by the processes. This is formalised by the following LTS for
monitors:
\myformulaM{$
\pp ? \set{\ell_i (\ST_i).\M_i}_{i \in I} \stackred{\pp ? \ell_j} \M_j ~~
j\in I\qquad\qquad
\sendMg{\pset}{S} \stackred{{\pset ! \ell_j}} \M_j ~~
j\in I\qquad\qquad
\myrule{\M\stackred{\alpha}\M'}{\M\pc\MM''\stackred{\alpha}\M'\pc\MM''}{}
$}
where $\alpha$ ranges over  $\pp?l$ and $\q!l$.

Rule \rulename{Com} allows monitored processes to exchange messages:
\myformula{$\myrule{\MM_1 \stackred{\q ? \ell } \MM_1' ~~~~~
\PP_1 \stackred{\ch[\pp] ? \ell (\val)} \PP_1' \quad\MM_2 \stackred{{\pp ! \ell}} \MM_2' ~~~~
\PP_2 \stackred{{\ch[\q] ! \ell(\val)}}  \PP_2'}
{\MM_1[\PP_1] \pcn  \MM_2[\PP_2]\red{}   \MM'_1[\PP'_1] \pcn  \MM'_2[\PP'_2]}{Com}
 $}
 Monitored processes can modify control data thanks to rule \rulename{OP}:
 \myformula{$\myrule{\PP \stackred{\put}\PP'}
 {\MM[\PP] \sig \sigma \red{} \MM[\PP'] \sig \put(\sigma)}{OP}$}

Adaptation is triggered by a function which 
applied to the state $\sigma$ 
returns a global type $\GG$ which represents the choreography of the new part of the system, a set $\KK$ of participants that are removed and a new state $\sigma'$.  The outcome of the adaptation function determines a reconfiguration of the system as summarised below.
Let $\HH$ denote the set of the participants before adaptation.
The system running before the adaptation step is modified in the following way:
    \begin{enumerate}
 \item the participants in $\KK$ are removed;
 \item all communications between participants in $\HH\setminus\KK$ and participants in $\KK$ are erased from all monitors of participants in $\HH\setminus\KK$;
 \item\label{1c} all communications between two participants in $\HH \cap\pt(\GG)$ are erased from all monitors of participants in $\HH \cap \pt(\GG)$.
 \item The monitors resulting as projections of $\GG$ are added, in parallel with the monitors of  participants in $\HH \cap \pt(\GG)$ obtained as described in~(\ref{1c}), or as monitors of new participants in $\pt(\GG)\setminus\HH$.
 \item The processes with adapted monitors are modified in order to be adequate to these monitors.
\end{enumerate}

  To define the formal rule for adaptation we need to define the mappings $\mon$ and $\proc$.
By $\MM\setminus\HH$ we denote the monitor obtained from $\MM$  by erasing all communications to  participants belonging to $\HH$.
\myformula{$\mon(\pp,\MM,\GG,\KK)=\begin{cases}
 \MM\setminus\KK      & \text{if }\pp\not\in\pt(\GG), \\
  (\proj\GG\pp)\pc(\MM\setminus(\pt(\GG)\cup\KK))   & \text{otherwise}.
\end{cases}$}
The mapping $\mon$ applied to a participant $\pp$, his current monitor $\MM$, a global type $\GG$ and a set of killed participants $\KK$ gives the new monitor for $\pp$. If $\pp$ is not a participant of $\GG$, then the new monitor is simply $\MM$ where all communications with killed participants are erased.
Note that adaptation is transparent to $\pp$ whenever $\pp$ does not communicate with killed participants. Otherwise the new monitor is the parallel composition of the projection of $\GG$ onto $\pp$ with the monitor $\MM$, where all communications with participants of $\GG$ and killed participants are erased.
Clearly $\mon$ is a partial mapping, since it is undefined when some labels occur both in $\proj\GG\pp$ and in $\MM\setminus(\pt(\GG)\cup\KK)$. We could make $\mon$ total simply by always performing a fresh renaming of the labels in $\GG$. This dramatic solution however would never allow  us to use the current process for a participant in $\pt(\GG)$, even if the process is adequate for the new monitor. 
For this reason in rule \rulename{Adapt}
by $\hat\GG$ we denote $\GG$ if $\mon(\pp,\MM,\GG,\KK)$ is defined for all $\pt(\GG)$ and a fresh relabelling of $\GG$ otherwise.
\myformula{$\procS(\ch[\pp],\P,\M,\HH)=\begin{cases}
  \P    & \text{if }\P\agree\M\setminus\HH,\\
 \Q\sub{\ch[\pp]}y& \text{if }\P\not\agree\M\setminus\HH~\&~\Q\in\procset~\&~ \Q\agree \MM\setminus\HH.
\end{cases}$}
The mapping $\procS$ deals with a single threaded process $\P$ (with channel $\ch[\pp]$), a single threaded monitor $\M$ and a set $\HH$ of processes. If $\P$ is adequate for $\M$ without the communications to participants in $\HH$, i.e. for $\M\setminus\HH$, then the mapping returns $\P$. Otherwise the mapping takes a process $\Q$ (adequate for $\M\setminus\HH$) in the collection $\procset$ and it returns $\Q$ where $y$ is replaced by $\ch[\pp]$, i.e. $\Q\sub{\ch[\pp]}y$.
\myformula{$\begin{array}{l}\proc(\ch[\pp],\PP,\MM,\GG,\KK)=\begin{cases}
  \PP    & \text{if }\PP\agree\mon(\pp,\MM,\GG,\KK)\\
  \Pi_{i\in I}\procS(\ch[\pp],\PP.i,\MM.i,\KK)     & \text{if }\pp\not\in\pt(\GG)\\
 \QQ\sub{\ch[\pp]}y\pc\PP'     & \text{if }\pp\in\pt(\GG)~\&~\QQ\in\procset~\&~\QQ\agree \proj\GG\pp
\end{cases}\\
\hfill\text{where }\PP'=\Pi_{i\in I}\procS(\ch[\pp],\PP.i,\MM.i,\pt(\GG)\cup\KK).\end{array}$}
The mapping $\proc$ returns $\PP$ whenever $\PP$ is adequate for $\mon(\pp,\MM,\GG,\KK)$. Otherwise both $\PP$ and $\MM$ are split in the corresponding single threaded processes ($\PP.i$) and monitors ($\MM.i$) for $i\in I$ according to Lemma~\ref{la}. Each $\PP.i$ and $\MM.i$ for $i\in I$ become arguments of $\procS$. If $\pp\not\in\pt(\GG)$ the set of processes argument of $\procS$ is $\KK$. Otherwise the set of processes argument of  $\procS$ is $\pt(\GG)\cup\KK$. This reflects the two cases in the definition of $\mon$. Moreover if $\pp\in\pt(\GG)$ the final parallel contains also a process $\QQ$ in $\procset$ (adequate for $\proj\GG\pp$) with $y$ replaced by  $\ch[\pp]$.\\
For example $\proc(\ch[\pp],\PP,\MM,\GG,\set{\pr})=\ch[\pp]?\ell_1(x)\pc\ch[\pp]!\ell_4(7)$ if $\PP=\ch[\pp]?\ell_1(x)\pc\ch[\pp]?\ell_2(y)\pc\ch[\pp]!\ell_3(\true)$,
$\MM=\q?\ell_1(\Nat)\pc\pr?\ell_2(\Nat)\pc\q'!\ell_3(\Bool)$, $\GG=\pp\to\q':\ell_4(\Nat)$, and  $\procset$ contains $y!\ell_4(7)$.

We can now define the adaptation rule:
\myformula{
$$\myrule {\begin{array}{c}\fun(\sigma) = (\GG,\KK,\sigma')\quad\forall\pp\in\pt(\GG)\setminus\HH.\QQ_\pp\in\procset~\&~ \QQ_\pp\agree \proj{\hat\GG}\pp\\
    \MM_\pp'=\mon(\pp,\MM_\pp,\hat\GG,\KK)\quad\PP_\pp'=\proc(\ch[\pp],\PP_\pp,\MM_\pp,\hat\GG,\KK)\end{array}}
 { \rest{\ch}~ (\prod_{\pp \in \HH} \MM_\pp [\PP_\pp])  \sig \sigma \red{}\rest{\ch}~ (\prod_{\pp \in \HH\setminus\KK} \MM_\pp' [\PP_\pp'])\pcn\prod_{\pp \in \pt(\GG)\setminus\HH} \proj{\hat\GG}\pp [\QQ_\pp\sub{\ch[\pp]}y])  \sig \sigma'}{Adapt}
 }
Rule \rulename{Adapt} must be used when the adaptation function $\fun$ applied to the control data $\sigma$ is defined. In this case $\fun$ returns a global type $\GG$, a set of participants $\KK$ which must be killed and a new control data $\sigma'$. The global type $\hat\GG$ is $\GG$ with possibly some relabelling, as defined by rule $\rulename{adapt}$.
The set of participants before the adaptation is $\HH$. After the adaptation the set of participants is $(\HH\setminus\KK) \cup (\pt(\GG)\setminus\HH)$, i.e. the initial set of participants without the killed ones, plus the participants of $\GG$ which do not occur in $\HH$. For these new participants the processes are taken from the collection $\procset$ as in rule \rulename{InitS}. Instead for the participants in $\HH\setminus\KK$ we compute the new monitors and processes using the mappings $\mon$ and $\proc$, respectively.\\


Table~\ref{sr} lists the reduction rules of networks and systems, which are not discussed above.
Evaluation contexts are defined by \myformulaH{$\E::= [\;] ~\sep~ \E\pcn \N ~\sep~ (\nu s) \E$.}
\begin{table}[b]\normalsize
\centerline{$\begin{array}{c}
\myrule{\PP \stackred{\tau}  \PP'}{\MM[\PP]\red\MM[\PP']}{Tau}
 \qquad
 \myrule{\N_1\equiv \N_1' \quad \N_1'\red \N_2' \quad \N_2\equiv \N_2'}{\N_1\red \N_2}{Equiv}\\[20pt]
 \myrule{\N\red{}\N'}{\E[\N]\sig\sigma\red\E[\N']\sig\sigma}{SN}\qquad \myrule{\N\sig\sigma\red{}\N'\sig\sigma'}{\E[\N]\sig\sigma\red\E[\N']\sig\sigma'}{CTX}
 \end{array}$}
 \caption{\normalsize Further network and system reductions.}\labelx{sr}
\end{table}

 A system starts reducing session initiators.  Then, in each reduction step, rule \rulename{Adapt} is used whenever the adaptation function applied to the control data is defined. Otherwise the application of rule  \rulename{OP} has precedence over the remaining rules for each monitored process.

 A system is \emph{initial} if it is the parallel compositions of $\new$s.
 Starting from an initial system the reduction rules produce only monitored processes $\MM[\PP]$ which satisfy $\PP\agree\MM$. The crucial case is rule
\rulename{Adapt}, where the mapping $\proc$ builds the processes so that the adequacy between processes and monitors can be guaranteed.
\begin{mytheorem}
If $ \NS$ is an initial system  and $ \NS\red{}\NS'$, then all monitored processes in $ \NS'$ satisfy the adequacy condition.
\end{mytheorem}
The proof of this theorem requires typing rules for networks and does not fit in this paper.


The closure $\CC(\Sigma)$ of a set $\Sigma$ of monitors is the smallest set which contains $\Sigma$ and such that:
\begin{myitemize}
\item $\M\in\CC(\Sigma)$ implies $\M\setminus\KK\in\CC(\Sigma)$ for an arbitrary $\KK$;
\item $\MM\in\CC(\Sigma)$ and $\MM'\in\CC(\Sigma)$ imply $\MM\pc\MM'\in\CC(\Sigma)$.
\end{myitemize}

We can also guarantee progress of systems under a natural condition on the collection $\procset$. A collection $\procset$ is \emph{complete for a given set of global types} $\mathcal{G}$ if, for every $\GG\in\mathcal{G}$, there are processes in $\procset$ for the closure of the set of monitors obtained from $\GG$ by projecting onto participants.

\begin{mytheorem}
If $\procset$ is complete for $\mathcal{G}$, the adaptation function $F$ only produces global types in $\mathcal{G}$ and $ \NS$ is an initial system with global types in $\mathcal{G}$, then $ \NS$ has progress.
\end{mytheorem}
The proof of this theorem uses previous theorem and the technique illustrated in~\cite{CDYP2015}.

\mysection{Conclusion and Related Works}\label{crw} To the best of our knowledge the present paper provides the first formal model of self-adapting multiparty sessions in which control data trigger the system reconfigurations.

 As already mentioned, our work builds on~\cite{CDV15}, where a calculus based on global types, monitors and processes similar to
  the present one was introduced. There are two main points of departure from
  that work. First, in the calculus of~\cite{CDV15},  the global type decides \emph{when} the adaptation can take place, since its monitors prescribe when some participants have to check data and then send a request of adaptation to the other participants. Second, in~\cite{CDV15} any adaptation involves all session participants, by requiring a global new reconfiguration of the whole system.
In contrast, in our calculus adaptation is only triggered  by dynamic changes in control data,  so giving rise to unpredictable reconfiguration steps. Furthermore, reconfiguration can involve a partial set of communications only, when a 
part of the behaviour must be changed. The main technical novelty allowing these features is the parallel composition of monitors.
Moreover, communication is synchronous here and asynchronous in~\cite{CDV15}. We adopted this choice with the goal of designing a calculus as simple as possible, while capturing main issues concerning a safe self-reconfiguration of the participant behaviours in unuspected adaptations. Using asynchronous communications would add the technical  complication of reconfiguring  channel queues during adaptation, which is left for future work.

Works addressing adaptation for multiparty communications include~\cite{AR12},~\cite{DBLP:conf/sefm/BravettiCHLMPZ13}, ~\cite{DBLP:conf/sle/PredaGLMG14}
  and~\cite{CDP14}.  In paper~\cite{AR12} global and session types are used to guarantee deadlock-freedom in a calculus of multiparty sessions with asynchronous communications. Only part of the running code is updated. Two different conditions are given for ensuring liveness. The first condition requires that all channel queues are empty before updating. The second condition requires a partial order between the session participants with a unique minimal element. The participants are then updated following this order. We do not need such conditions for progress, since communication is synchronous and adaptation is done in one single big step.
  The work~\cite{DBLP:conf/sefm/BravettiCHLMPZ13}
  enhances a choreographic language with constructs defining
  adaptation scopes and dynamic code update; an associated endpoint
  language for local descriptions, and a projection mechanism for
  obtaining (low-level) endpoint specifications from (high-level)
  choreographies are also described. A typing discipline for these
  languages is left for future work.  The
  paper~\cite{DBLP:conf/sle/PredaGLMG14} proposes a choreographic
  language for distributed applications.  Adaptation follows a
  rule-based approach, in which all interactions, under all possible
  changes produced by the adaptation rules, proceed as prescribed by
  an abstract model. In particular, the system is deadlock-free by
  construction. The adaptive system is composed by interacting
  participants deployed on different locations, each executing its own
  code. The calculus of~\cite{CDP14} is inspired by~\cite{CDV15}, the main difference being that security violations trigger adaptation mechanisms that prevent violations to occur and/or to propagate their effect in the systems.

  There is substantial difference   between our monitors and run-time enforcement monitors of~\cite{S00},  where monitors terminate the execution if it is about to violate the security policy being enforced.



Future work includes the addition of a local state to each participant. This can allow us to model internal interactions, based on local data, among single threaded processes whose parallel composition implements a participant.
Using both internal interactions and communications among participants enables the application of our calculus to more realistic cases of study.
In general, we plan to consider application-driven instances of our
global adaptation mechanisms, in which concrete adaptation functions are used.

Finally, we intend to investigate whether our approach can be useful in providing a formal model to mechanisms of code hot-swapping in programming languages (such as Erlang).

\myparagraph{Acknowledgments}
We are grateful to the anonymous reviewers for their useful suggestions and remarks.

\vspace{-10pt}
\bibliographystyle{eptcs}
\bibliography{adapt}



\end{document}